\documentclass[12pt]{article}
\usepackage{a4wide}
\usepackage{graphicx}% Include figure files

\advance\oddsidemargin -3mm
\advance\evensidemargin -3mm

\newcommand{\ms}{m_{\tilde \tau}}
\newcommand{\st}{{\tilde \tau}}
\newcommand{\ET}{E_{\rm T}}
\newcommand{\mET}{\mbox{$\slash \!\!\!\! \ET$}}
\newcommand{\pT}{p_{\rm T}}
\newcommand{\Njet}{N_{\rm jet}}
\newcommand{\HT}{H_{\rm T}}
\newcommand{\ST}{S_{\rm T}}
\newcommand{\tauh}{\tau_{\rm h}}

\newcommand{\beq}{\begin{equation}}
\newcommand{\eeq}{\end{equation}}
\newcommand{\beqa}{\begin{eqnarray}}
\newcommand{\eeqa}{\end{eqnarray}}
\newcommand{\beqar}{\begin{eqnarray*}}
\newcommand{\eeqar}{\end{eqnarray*}}
\newcommand{\eg}{{e.g.,}\ }
\newcommand{\ie}{{i.e.,}\ }
\begin{document}
\thispagestyle{empty}

\hfill{\sc UG-FT-297/12}

\vspace*{-2mm}
\hfill{\sc CAFPE-167/12}

\vspace{32pt}
\begin{center}
\centerline{\textbf{\Large
Supersymmetry with long-lived staus
at the LHC}}

\vspace{40pt}

R.~Barcel\'o, J.I.~Illana, M.~Masip, A.~Prado, and P.~S\'anchez-Puertas
\vspace{12pt}

\textit{
CAFPE and Departamento de F{\'\i}sica Te\'orica y del
Cosmos}\\ \textit{Universidad de Granada, E-18071, Granada, Spain}\\
\vspace{16pt}
\texttt{rbarcelo,jillana,masip,aprado,pablosp@ugr.es}
\end{center}

\vspace{40pt}

\date{\today}% It is always \today, today,
             %  but any date may be explicitly specified

\begin{abstract}

We consider SUSY extensions of the standard model 
where the gravitino 
is the dark-matter particle and the stau is long lived.
If there is a significant mass gap with 
squarks and gluinos, the staus produced at 
hadron colliders tend to be fast ($\beta>0.8$), and 
the searches based on their delay in the time of flight
or their anomalous ionization become less effective. Such 
staus would be identified as regular muons with the 
same linear momentum and a slightly reduced energy.
Compared to the usual SUSY models where a neutralino
is the LSP, this scenario implies {\it (i)}
more leptons (the two staus at the end of the decay chains), 
{\it (ii)} a strong
$e$--$\mu$ asymmetry, and {\it (iii)} 
less missing $\ET$ (just from neutrinos, 
as the lightest neutralino decays into stau).
We study the bounds on {\it this} SUSY from current
LHC analyses (same-sign dileptons and  
multilepton events) and discuss the 
best strategy for its observation.

\end{abstract}

\newpage

\section{Introduction}

The determination of the mass and the couplings of
the Higgs boson at the
LHC will not {\it complete} our understanding of the 
mechanism responsible 
for the breaking of the electroweak (EW) symmetry. 
It will be also essential to establish whether or
not there is a dynamical principle explaining its
nature. Supersymmetry (SUSY) 
is a possibility that has attracted a lot of work
during the past decades. Minimal SUSY extensions with 
a neutralino as the lightest SUSY particle (LSP) provide a
good candidate for dark matter, imply a consistent 
picture for gauge unification, and can in principle 
accommodate a 125 GeV light Higgs \cite{Heinemeyer:2011aa}.
It is apparent that 
the non-observation of flavor-changing neutral currents 
or electron and neutron electric dipole moments requires
an {\it effort} in these frameworks. However, SUSY models
have proven flexible enough to adapt, and they have
reached the current phase of direct search at the LHC in a 
(reasonable) good shape.

SUSY searches at hadron colliders have focused on a few generic
signals with relatively small backgrounds. The {\it classic}
one \cite{Gamberini:1986eg} is jets with no hard 
leptons but large $\mET$ from squarks $\tilde q$
going into a quark $q$ plus the lightest neutralino 
$\tilde\chi_1^0$. 
It was then emphasized \cite{Barnett:1993ea} that chain decays
of colored SUSY particles 
through charginos and heavier neutralinos giving two isolated
leptons usually have a much larger branching ratio. 
In particular, gluino pairs  
provide same-sign (SS) dileptons together with jets
and $\mET$, a clean signal of high discovery 
potential \cite{Kraml:2005kb}.
Initial searches at the 7 TeV LHC do not show any hints of
such signals and set bounds on squark and gluino masses
that rise up to 800 GeV and higher, although a complete
exclusion of this mass region in the neutralino LSP model would
require a careful consideration of some cases with an anomalous
signal \cite{Fan:2011yu,Csaki:2012fh,Drees:2012dd}.

There are, however, other SUSY scenarios that provide a
{\it different} generic signal, and one may wonder how 
constrained they are by current LHC analyses.
In particular, 
a possibility that is well motivated from a 
model-building point of view is the case with a gravitino LSP. 
This is natural in all models
with a low scale of SUSY breaking, like the ones mediated by
gauge interactions \cite{Giudice:1998bp,Arcadi:2011yw}. Even in 
gravity-mediated models, the LSP gravitino may be an acceptable
dark matter candidate with \cite{Buchmuller:2007ui} 
or without \cite{Ellis:2003dn} $R$-parity violation.
In all these cases the next-to-LSP could be a long-lived 
charged particle (\eg  
the $\tilde \tau$) that, if produced at the LHC, 
would decay after crossing the detectors. 

The search strategy in these scenarios 
is then different \cite{Ellis:2006vu}. A charged 
particle of mass $\ms$ and three-momentum $p=\beta\gamma \ms$ 
will curve under the magnetic field in the inner detector like
a muon of the same momentum. There are, however, two observables
that could  distinguish such a heavy muon:
an anomalous ionization in the silicon tracking detector and
a delay in the time of flight from the vertex to the muon chambers.

\begin{figure}[t]
\begin{center}
\begin{tabular}{cc}
\includegraphics[width=0.48\linewidth]{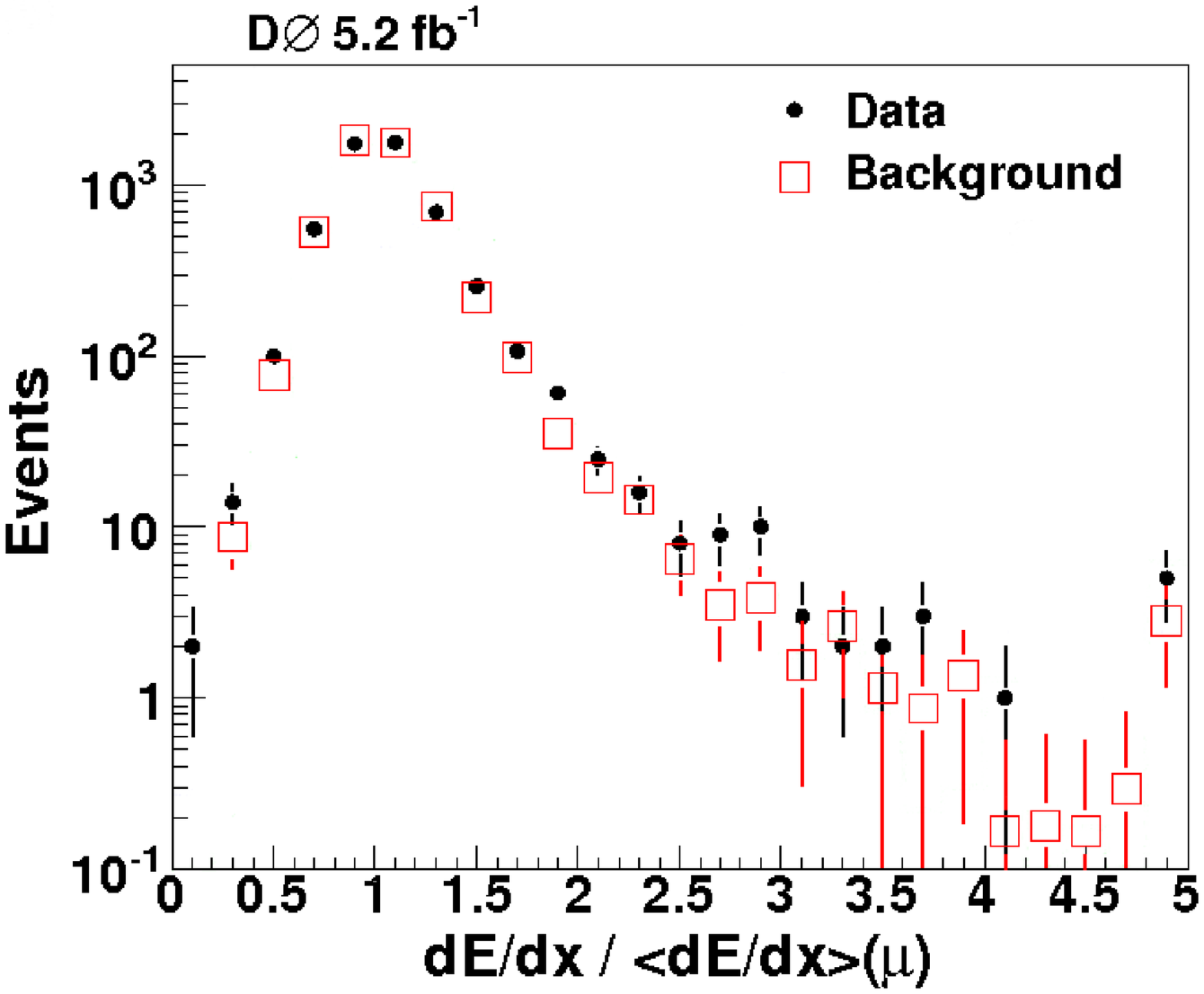}
\includegraphics[width=0.48\linewidth]{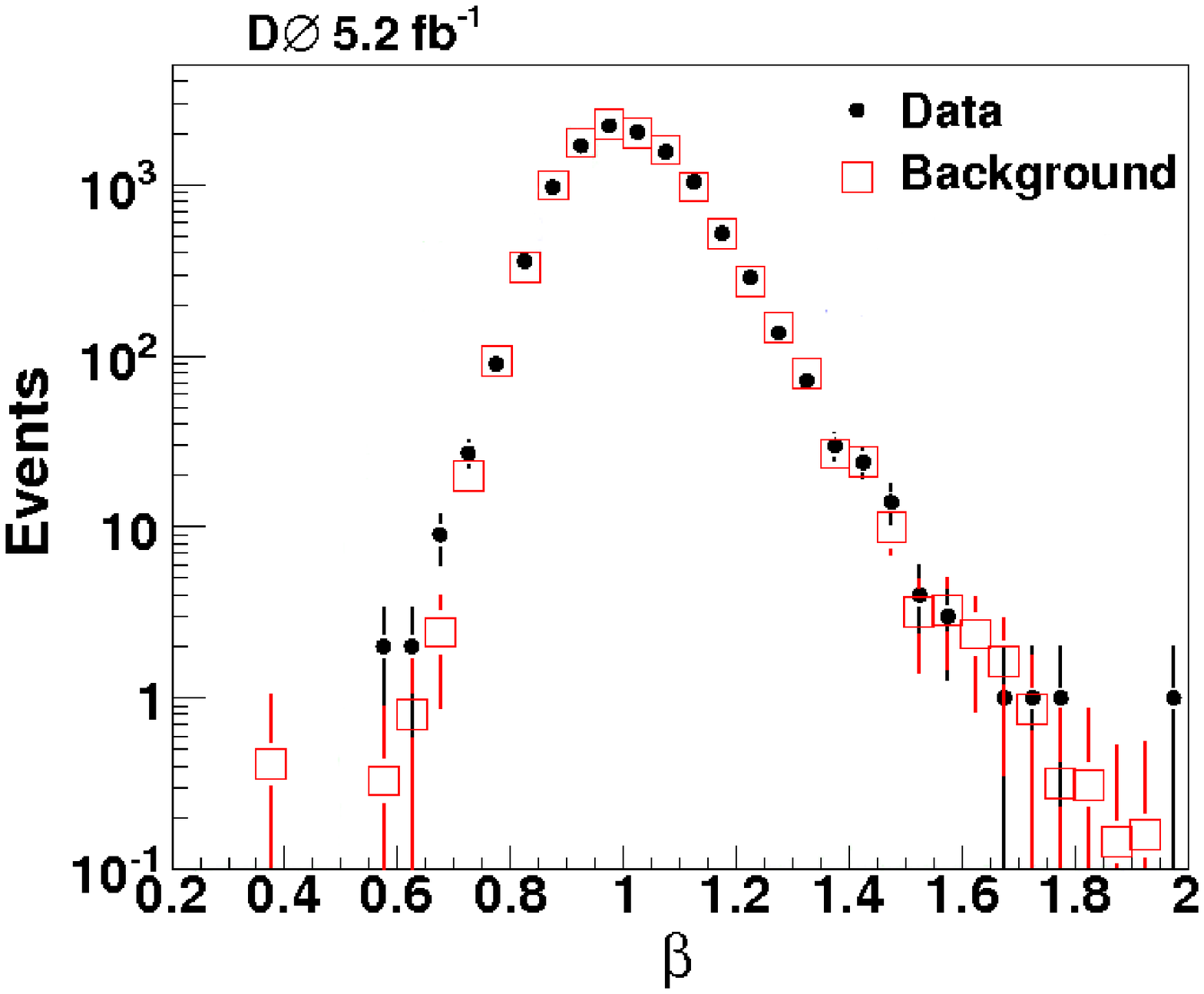} 
\end{tabular}
\end{center}
\caption{Distributions of ${\rm d}E/{\rm d}x$ (left) and speed $\beta$ (right)
observed at D0 (from \cite{Abazov:2011pf}). 
The scale of ${\rm d}E/{\rm d}x$ is adjusted so that the 
distribution from $Z\to \mu\mu$ peaks at 1.}
\label{CDF}
\end{figure}

As a stau or a muon propagate in matter, 
low $q^2$ processes 
like ionization are insensitive to the mass, and one expects
that the effects on the medium will only depend on the velocity 
(or $\beta \gamma$) of the particle. The Landau most probable energy 
deposition through ionization is large 
at low values of $\beta$ (it goes like $( \beta\gamma )^{-2}$), 
has a minimum at 
$\beta \gamma\approx 4$ and reaches the so called Fermi plateau
at $\beta \gamma > 100$ (see Fig.~30.9 at the PDG \cite{Nakamura:2010zzi}). 
In particular, the ionization along the track
of a 100 GeV stau of $\beta\gamma=2$ (\ie
$\beta=0.89$ and $p=200$ GeV) would be very similar to that of 
a muon of the same momentum,
and 25\% higher at $\beta \gamma = 1$ 
(or $\beta=0.7$). In Fig.~\ref{CDF}--left we reproduce a 
plot from the Tevatron D0 experiment
\cite{Abazov:2011pf} of
${\rm d}E/{\rm d}x$ relative to the average value for
muons passing certain 
$\pT$, rapidity and isolation cuts.
For an actual stau, given the width of the expected distribution 
(around $30\%$ of its average value, 
see Fig.~30.8 at the PDG \cite{Nakamura:2010zzi})
and the uncertainty in the response of the detector,
one could expect a clear difference with
muons only for $\beta\le 0.7$.

The direct measure of $\beta$ has just a slightly better
resolution. At D0 (see Fig.~\ref{CDF}--right) 27\% of the 
muons are measured with $\beta>1.1$, and 3.5\% of the 
{\it subluminal} ones have $\beta<0.8$. A 37 pb$^{-1}$
ATLAS analysis \cite{Aad:2011hz} at
the 7 TeV LHC
shows a more accurate description of the muon
velocity, setting the limit $\ms>110$ GeV from direct stau
production. A recent study \cite{Chatrchyan:2012sp} by CMS using  
5.0 fb$^{-1}$ of data could imply higher bounds. It is 
difficult, however, to use
their results to constrain a particular model, since 
{\it (i)} they do not provide the complete velocity
distribution observed for muons, including the region
with $\beta>1$ (necessary to estimate the effect of
the reconstruction on the stau velocity), and 
{\it (ii)}
they could be overestimating the anomalous 
ionization of heavy particles. In particular, their 
method seems to imply 
a 10\% excess for a stau of $\beta=0.9$, when such particle
is below the Fermi plateau and should
ionize like a muon of the same 
three-momentum (see \cite{Chen:2009gu}
for a discrimination based on radiative energy deposition).

\begin{figure}[t]
\begin{center}
\begin{tabular}{cc}
\includegraphics[width=0.5\linewidth]{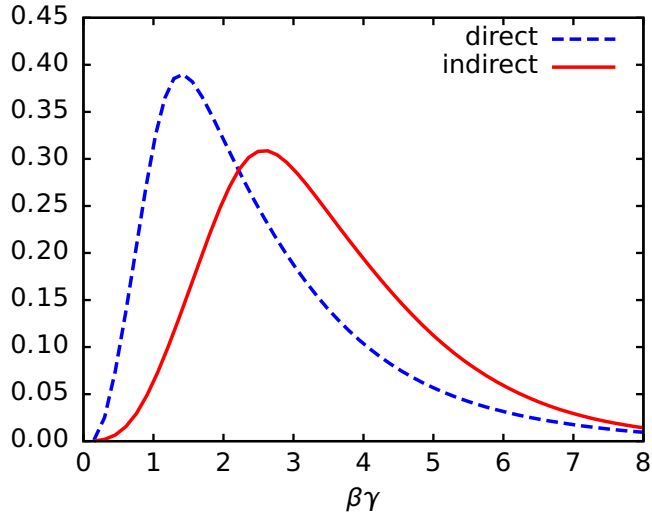} 
\end{tabular}
\end{center}
\caption{$\beta\gamma$ distribution (normalized to 1) of 150 GeV staus 
from direct production and from chain decays of
800 GeV squarks and 1 TeV gluinos (direct production accounts for 
1 out of 15 staus produced).}
\label{beta}
\end{figure}
In addition, in models where the stau is significantly
lighter than squarks and gluinos its velocity tends to 
be high (see \cite{Heisig:2012zq}
for an analysis of the kinematics in these 
chain decays), and one is left with relatively 
few events with a $\beta$ small enough to give a
clear deviation in the two observables. 
Let us take, just 
for illustration, a 150 GeV  
$\tilde \tau_R$ together with 750 GeV Higgsinos, 800 GeV
light-flavor squarks and 1 TeV gluinos, with the 
slepton doublets and the rest of squarks and gauginos 
in the 800--1000 GeV mass region.
We will take the other two $\tilde \ell_R$ sleptons with 
a mass similar to $m_{\tilde \tau_R}$ (see next section).
The cross section for direct (Drell-Yan) 
production at the 7 TeV LHC is
around 33 fb, which is reduced to 31 fb 
once we require at least one
stau with $\pT>40$ GeV and  $| \eta | < 2.5$. 
In contrast, indirect production through squarks and gluinos
gives $\sigma=429$ fb, or $\sigma=428$ fb once we impose 
the same $\pT$ and rapidity requirements. This accounts for
14 times more stau pairs from indirect than from direct 
production. We plot
their $\beta\gamma$ distribution in Fig.~\ref{beta}.
While $28\%$ of the staus from direct production
have $\beta<0.8$, only $5\%$  of the ones 
from chain decays are in this $\beta$ region. If we restrict to
the $\beta<0.7$ (where a more significant anomaly can
be expected) these percentages are reduced to $14\%$ and 
$1\%$, respectively.

Therefore, in these models most of the 
events will contain two staus that look like regular muons
of momentum $\vec p_\mu=\vec p_\st$ and energy 
$E_\mu=\sqrt{E_\st^2-m_\st^2}$. Although specific analyses
have been proposed \cite{Galon:2011ws}, one may also ask how the usual 
SUSY searches constrain these scenarios assuming that the staus 
are identified as muons, and 
how to modify the cuts in order to optimize the
search. In this paper we study the bounds from recent studies on 
SS-dilepton \cite{ATLAS:2012ai} and inclusive-multilepton 
\cite{Chatrchyan:2012ye} production at the LHC.

\section{Same-sign leptons, jets and $\mET$}

SS dileptons can be an important signature in neutralino LSP 
models when gluinos are at accessible energies. 
If the collision produces $\tilde g \tilde g$ pairs that 
decay into charginos and neutralinos other than $\chi_1^0$, 
SS leptons will be very frequent, as each decay chain can
give a lepton or an antilepton with equal probability.
In addition, gluinos must decay into (real or
virtual) squarks producing jets, and there will also be $\mET$
from the undetected neutralino LSP. The same type of 
signal (with a smaller number of jets) 
may also be obtained from $\tilde u \tilde u$ pairs produced
through gluino exchange in the $t$--channel.

\noindent{\bf Gluinos in neutralino LSP models.}
In a recent (2.05 fb$^{-1}$ at 7 TeV) study \cite{ATLAS:2012ai}
ATLAS selects events in which the two higher-$\pT$
leptons ($\ell=e,\,\mu$) have the same charge, 
with at least 4 jets of
$\pT>50$ GeV, and with $\mET > 150$ GeV (plus certain isolation and
rapidity cuts). They estimate a background of about 1 event
from $t\bar t X$, fake leptons ($b$ or $c$-hadron decays), 
charge misidentification and dibosons, while they 
observe no events in the data.

Then this result is used to constrain
the signal from 650 GeV gluinos that decay
into $t \bar t \tilde\chi^0_1$ through a virtual stop of 1.2 TeV. 
They assume a 150 GeV neutralino and search for the channel
where two of the four final top
quarks give SS leptons.
 They predict around 7 events satisfying all the
requirements, which allows them to exclude the model.
We have reproduced their study in order to 
understand the differences with the long-lived stau
(LLST) scenario. In our analysis we have used {\tt MadGraph~5}
\cite{Alwall:2011uj} to obtain the $\tilde g\tilde g$ and 
the $\tilde g\tilde g+{\rm jet}$
cross sections, {\tt Prospino~2.1} \cite{Beenakker:1996ed}
to estimate next-to-leading 
order corrections, 
{\tt PYTHIA~6.4} \cite{pythia} for hadronization/showering effects 
and {\tt PGS~4} \cite{PGS4} (tuned to ATLAS in this study
and to CMS in the multilepton analysis) for detector simulation. 

We find that at the given luminosity 
a 650 GeV gluino mass implies the production of 1047
$\tilde g \tilde g$ pairs. A factor of $\epsilon=0.55$
must be included to take into account
the detector reconstruction, identification
and trigger efficiency, leaving the number of observable
pairs in ${\cal L} \sigma \epsilon=576$. 
The detection of two SS leptons (from $t$ decays) is 
then a very
selective requirement, reducing the 
signal to just 18 expected events. The successive 
cuts $\Njet>3$ and $\mET > 150$ GeV 
reduce this number further to 12 
and 7 events, respectively. Although these two cuts do not affect 
significantly the signal, they are essential to reduce the
background. The total acceptance after cuts
is $A=1.2\%$, implying a visible cross section 
$\sigma_{\rm vis}=\sigma \epsilon A=3.2$ fb that is above 
the $\sigma_{\rm vis}<1.6$ fb limit 
established by ATLAS.

\noindent{\bf Gluinos in long-lived stau models.}
Generically, the LLST scenario will imply a signal with 
some basic differences versus the neutralino LSP case:
\begin{itemize}
\item Two extra {\it leptons}, as SUSY particles are produced in
pairs and each one will chain-decay into a stau.

\item A strong $\mu$--$e$ asymmetry, as these staus taken
for leptons look always like muons.

\item Less $\mET$, as the lightest neutralino does not escape 
detection but decays into visible $\ell\tilde \ell$ pairs. 
\end{itemize}
Some comments about the second and third points above, however, 
are here in order. To be definite we will 
take $\st_1\approx \st_R$, and
$\tilde e_R, \tilde \mu_R$ of similar mass (as suggested by flavor
and other precision observables). This means that, depending
on the degree of degeneracy, when a 
$\tilde e$ is produced it may or may not decay into a $\st$ 
inside the detector 
(\eg $\tilde e\to \st e \tau, \st \nu_e \nu_\tau, \dots$). 
If $\tilde e$
escapes without decaying, it will just look like a 
long-lived stau. If $\tilde e$ decays promptly
(we neglect the possibility with displaced vertices) 
the resulting 
$\st$ will take a very large fraction 
of the selectron energy, and none of 
the extra particles (charged leptons and/or photons) 
will have enough $\pT$ to pass the cuts.
Moreover, since the $\tilde e$ boost is not ultrarelativistic
(typically $E_{\tilde e}/m_{\tilde e}=2$--$5$), the extra particles
will not be very focused along the stau direction and will not affect 
substantially its isolation cuts. Therefore, we can consider that 
the three $\tilde \ell_R$ {\it are} effectively long-lived staus
looking like muons.
Regarding the amount of $\mET$ in this scenario, 
notice that if the last step in the decay is not 
$\tilde\chi^0\to \ell^\pm \tilde \ell^\mp_1$ but 
$\tilde\chi^\pm_1\to \tilde \ell^\pm_1 \nu_\ell$, 
then $\mET$ will not be necessarily small. In
particular, if $m_{\tilde\chi^\pm}\gg \ms$ then the final neutrino will
take close to half of the chargino energy.

Let us then perform the ATLAS analysis assuming the LLST 
scenario. We will start with the case with a 150 GeV stau 
(together with $\tilde e_R$, $\tilde \mu_R$  
of similar mass) instead of the neutralino 
$\tilde\chi_1^0$ (assumed to be mostly a Bino), 
which is moved to 200 GeV. In our simulation we will  
just {\it change} these three sleptons ($\tilde \ell^\pm_1$)
to muons of the same 
three-momentum. The 650 GeV gluinos, like in their study, will
be forced to decay through a virtual stop into the neutralino,
\eg
\beq
\tilde g \to \tilde t\, \bar t \to \tilde\chi^0_1\, t\, \bar t
\to \tilde \ell^\pm_1 \,\ell^\mp\, t \,\bar t\,.
\eeq
We find that the 576 gluino pairs yield 209 SS-dilepton events
after the geometric and kinematic cuts, with 185 of
them including at least 4 jets of $\pT>50$ GeV. 
The large acceptance reflects the presence of the extra 
lepton produced in our framework. The requirement 
$\mET > 150$ GeV, however, reduces the 185 events to just 
18, defining  a signal that is a bit larger than the one obtained 
in the neutralino LSP scenario. We obtain an acceptable model
if the gluino mass is increased to 890 GeV, with only 
3.5 events surviving the cuts on 50 initial gluino
pairs.

The possibility
that most signal events are cut by the $\mET$ requirement is 
frequent in these LLST models. For example, if the 650 GeV 
gluinos are forced to decay
through a virtual light-flavor squark $\tilde q$ (instead
of the stop), we find 207 SS dileptons, 163 of them with at least
four very energetic jets, but only 2.3 events 
with large $\mET$. This result is mildly
dependent on the neutralino mass. If $m_{\tilde\chi^0_1}$
grows from 200 to 400 GeV the number of SS dileptons 
does not change, 
but the $\Njet>3$ cut is significantly stronger (as the total 
energy that goes into jets is smaller) and reduces the sample 
to 127 instead of 163 events. 
A heavier neutralino implies that the charged
lepton from its decay tends to carry more energy. If it
is a $\tau$ decaying leptonically, the energy taken by neutrinos
will also be larger. The
$\mET$ cut is then weaker in this case: we obtain  
a total of 13 events, 
which are enough to exclude the model. 
Therefore, we
find that the analysis would not exclude 650 GeV gluinos
for $m_{\tilde\chi^0_1}=200$ GeV but would imply 
$m_{\tilde g}\ge 830$ GeV if $m_{\tilde\chi^0_1}=400$ GeV.

\begin{table}[ht]
\begin{center}
\begin{tabular}{ c || c | c | c | c }
$m_{\tilde g}=650\;$  $m_{\tilde \tau}=150$ & $\;$ Signal $\;$ &
SS dilept. & $\Njet>3$ & 
$\mET>150$ \\ \hline    \hline
$\mu =200$  (through $\tilde t$) &576&199&159&21 \\ \hline
$\mu =400$  (through $\tilde t$) &576&194&89&63 \\ \hline
$\mu =200$  (through $\tilde q$) &576&208&158&12  \\ \hline
$\mu =400$  (through $\tilde q$) &576&204&114&65 \\ \hline
\end{tabular}
\caption{Number of events after cuts from gluino production, 
with Higgsinos decaying into the final staus 
(all masses in GeV). \label{gluino}}
\end{center}
\end{table}
As explained above, when charginos appear in the
gluino chain decay these models include a larger fraction
of events passing the 
$\mET$ cuts, \eg
\beq
\tilde g \to \tilde t\, \bar t \to \tilde\chi^+_1\, b\, \bar t
\to \tilde \ell^+_1 \,\nu_\ell\, b \,\bar t\,.
\eeq
Let us consider the case where they
go into relatively light Higgsinos, $\mu=200,400$ GeV with 
$M_{1,2}=700$ GeV. The results
are summarized in Table~\ref{gluino}, where we have assumed the 
same detector efficiency as in the previous study. 
We see that the signal is 
stronger than the one in analogous neutralino LSP scenarios, 
specially for values of the chargino mass significantly larger
than $\ms$. If the Higgsino mass is 400 GeV we obtain 
that the ATLAS analysis implies $m_{\tilde g}\ge 980$ GeV.

\noindent{\bf Squarks in long-lived stau models.} Let us 
comment on the limits implied by this analysis
when gluinos are heavier and the collision only 
produces squarks. Notice that in 
the neutralino LSP scenario considered by 
ATLAS with the squarks decaying into $q\tilde\chi_1^0$ 
the signal would not include charged leptons. In our case, however,
each neutralino will go into a muon-like slepton plus a lepton,
providing a signal.
Actually, these events would look similar to the gluino pairs 
studied before but with 
two fewer jets (or top quarks in $\tilde t\tilde t^*$ production).
Notice also that in the LLST scenario an event with a pair of
light-flavor squarks  
will not pass the $\Njet>3$ requirement unless
the squarks are produced with extra jets (a process that is 
included in our simulation) and/or 
the final $\tau$ lepton decays hadronically but is untagged.

\begin{table}[ht]
\begin{center}
\begin{tabular}{ c || c | c | c | c }
$m_{\tilde t}=650\;$  $m_{\tilde \tau}=150$ & $\;$ Signal $\;$ & 
SS dilepton & $\Njet>3$ & 
$\mET>150$ \\ \hline    \hline
$M_1=200$  & 11  & 4 & 2 & 0.29 \\ \hline
$\mu=200$  & 11  & 3.4 & 1.7& 0.15  \\ \hline
\end{tabular}
\caption{Number of events after cuts from stop production,
with Binos or Higgsinos decaying into the final staus. \label{stop}}
\end{center}
\end{table}
For the analysis of stop-pair production (in Table~\ref{stop}), we
take $\tilde t_1$ (mostly $\tilde t_R$) at 650 GeV 
with  the rest of squarks decoupled.
The signal will include SS dileptons and 4 jets if, for example,
one of the tops decays hadronically and the other
one leptonically, which would provide also $\mET$. 
We find, however, that the requirement $\mET>150$ GeV is
too strong (it reduces the acceptance to just 2.6\%) and the model
can not be excluded by the current analysis. If the stop can 
decay both to
charginos and neutralinos, \eg
$\tilde t\to b \tilde\chi^+\to b \tilde \tau^+ \nu$
and $\tilde t^*\to \bar t \tilde\chi^0\to \bar b q \bar q' \tau^-
\tilde \tau^+$, the 
channel with the two tops going through chargino does not
contribute and the signal is even weaker (0.07 events pass 
the cuts on the initial 11 $\tilde t\tilde t$ pairs). We have
included also this case in Table~\ref{stop}.

\begin{table}[ht]
\begin{center}
\begin{tabular}{ c || c | c | c | c }
$m_{\tilde q}=650\;$  $m_{\tilde \tau}=150$ & $\;$ Signal $\;$ & 
SS dilepton & $\Njet>3$ & 
$\mET>150$ \\ \hline    \hline
$M_1=200$ & 672 & 258 & 52 & 1.5 \\ \hline
$\mu=200$ & 672 & 275 & 48 & 4.6 \\ \hline
\end{tabular}
\caption{Number of events after cuts from squark production, 
with Binos or Higgsinos decaying into the final staus. \label{squark}}
\end{center}
\end{table}
To illustrate the case with light-flavor squark production,  
we take the first two families of squarks (L and R)
with $m_{\tilde q}=650$ GeV together with 
$1.5$ TeV gluinos. We obtain a total of 672 $\tilde q\tilde q$
events (90\% from gluino in the $t$ or the $u$ channels), with 
274 of them including an additional jet. In
Table~\ref{squark} 
we summarize our results when the squarks are forced to decay
to neutralinos ($M_1=200$ GeV and $M_2,\mu=700$ GeV) or
can also decay into charginos ($\mu=200$ GeV, $M_{1,2}=700$ GeV).
We observe that
the $\Njet>3$ cut is now severe and, again, the $\mET$ requirement 
puts the first case well below the background. The second
case, with one squark giving a chargino ($\tilde\chi^+\to \nu \tilde \tau^+$)
and the other one a neutralino ($\tilde\chi^0\to \tau^-_h \tilde \tau^+$),
implies more $\mET$, and squarks masses below 770 GeV would be 
excluded by these ATLAS results.

\noindent{\bf Optimized SS-dilepton search.} The search for 
LLST SUSY based on SS dileptons could be optimized by slightly 
adapting the cuts. The same ATLAS cuts used in the neutralino 
LSP search are optimal only for gluino production 
with stop and charginos in its chain decay. In the rest of the cases
the missing $\ET$ cut must be relaxed. The requirement of 
4 very energetic jets is optimal in the search for gluino 
production, but it must be also relaxed to $\Njet\ge 2$ 
in squark searches. In that case the background (which
tends to be larger) can be reduced requiring for another hard
lepton that combined with any of the SS leptons is off 
the $Z$ mass shell. In all the cases the SS-dilepton
excess exhibits a large electron--muon asymmetry, as 
long-lived sleptons look always like muons. If the 
$\tilde \tau$'s 
are obtained from Higgsino decays we obtain no $ee$ pairs
and just 3--10\% of $e\mu$ events, with the rest of
them defined by two muon-like particles.
For staus from parent gauginos there is  
1\% of $ee$, 20--30\% of $e\mu$, and
70--80\% of $\mu\mu$ events.

\section{Inclusive multilepton search}

In a recent work \cite{Chatrchyan:2012ye}
CMS has searched for an anomalous production
of multilepton events at the 7 TeV LHC 
for an integrated luminosity of 4.98 fb$^{-1}$. 
Their analysis is very complete and model-independent, it
applies to any scenario with new particles producing leptons
and certainly to our LLST model. They use $\HT$, defined
as the scalar sum of the $\pT$ of all reconstructed jets,
and the analogous $\ST$ (which includes the leptons 
and missing $\ET$) to detect the presence of 
heavy physics. They classify 
in a systematic way all the possibilities: 4 or 3 leptons; $\mET$
above or below 50 GeV; lepton pairs around the $Z$ mass or
not; and low or high values of $\HT$ or $\ST$. 
Moreover, they separate events with 0, 1 or 2 tau leptons
decaying hadronically into a single track 
(one-prong $\tauh$ decays). Being heavier, the third lepton
family tends to be more sensitive to the new physics. This is also
the case in all SUSY models, where 
the Higgsinos couple to taus but not significantly 
to muons or electrons.
Events with heavy particles decaying into leptons would appear 
in one or another of the bins that they consider, and the 
estimated background (which includes double vector-boson,
$t\bar t$, or $t\bar t V$ production) 
is particularly small in the $4\ell$ channels.

In LLST SUSY any event has at least 
two charged leptons (the two staus)
at the end of the decay chains. If the staus are produced through
neutralino the proces will also include extra leptons, whereas
the $\tilde\chi^\pm\tilde\chi^0$ channel 
implies a neutrino (\ie missing $\ET$ instead
of $\ell^\pm$) and an excess of three-lepton events.
Notice that if the lighter neutralinos are mostly 
Higgsinos the muon-like slepton will come with a $\tau$,
while gauginos will imply 
the three lepton flavors with the same frequency.

Under this multilepton analysis the difference between gluino and 
squark events is not so strong as in the SS-dilepton
search, since the number of jets is not a discriminating 
observable. Instead, the mass difference 
between the colored particles ($\tilde g$ or $\tilde q$)
produced in the collision and the chargino/neutralino mass 
becomes critical.
It is easy to see that if this mass difference is large the
event will have energetic jets and a large value of $\HT$,
whereas if it is small most of the energy will go to the
leptons.

\begin{table}[t!] 
\begin{center}
{\footnotesize 
\begin{tabular}{c|c|c|c}
\hline
Selection&N$(\tauh)=0$&N$(\tauh)=1$&N$(\tauh)=2$ \\ \hline\hline
  &  obs (SM) {\bf NP} & obs (SM) {\bf NP} & obs (SM) {\bf NP} \\ \hline
4 Lepton results & \multicolumn{3}{c}{} \\ \hline
$\mET>50$, $\HT>200$, no Z & 0 (0.018$\pm$0.005) {\bf 6.4} 
& 0 (0.09$\pm$0.06) {\bf 17} & 0 (0.7$\pm$0.7) {\bf 5.8} \\ \hline
$\mET>50$, $\HT<200$, no Z & 1 (0.20$\pm$0.07) {\bf 0.1} &
3 (0.59$\pm$0.17) {\bf 0.1}  & 1 (1.5$\pm$0.6) {\bf 0.1}\\ \hline
$\mET<50$, $\HT>200$, no Z &0 (0.006$\pm$0.001) {\bf 8.5} &
0 (0.14$\pm$0.08) {\bf 12} &0 (0.25$\pm$0.07) {\bf 4.0} \\ \hline
$\mET<50$, $\HT<200$, no Z &1 (2.6$\pm$1.1) {\bf 0.0} & 
5 (3.9$\pm$1.2) {\bf 0.0}&17 (10.6$\pm$3.2) {\bf 0.1} \\ \hline
3 Lepton results & \multicolumn{3}{c}{} \\ \hline
$\mET>50$, $\HT>200$, no-OSSF &2 (1.5$\pm$0.5) {\bf 45} &
33 (30.4$\pm$9.7) {\bf 62}  &15 (13.5$\pm$2.6) {\bf 2.5} \\ \hline
$\mET>50$, $\HT<200$, no-OSSF & 7 (6.6$\pm$2.3) {\bf 0.0}  &
159 (143$\pm$37) {\bf 0.4} &82 (106$\pm$16) {\bf 0.0} \\ \hline
$\mET<50$, $\HT>200$, no-OSSF &1 (1.2$\pm$0.7) {\bf 27} &
16 (16.9$\pm$4.5) {\bf 31} & 18 (31.9$\pm$4.8) {\bf 0.6}\\ \hline
$\mET<50$, $\HT<200$, no-OSSF & 14 (11.7$\pm$3.6) {\bf 0.1} &
446 (356$\pm$55) {\bf 0.0} &1006 (1026$\pm$171) {\bf 0.0} \\ \hline
$\mET>50$, $\HT>200$, no Z &8 (5.0$\pm$1.3) {\bf 116}  &
16 (31.7$\pm$9.6) {\bf 62}  & -\\ \hline
$\mET>50$, $\HT<200$, no Z & 30 (27.0$\pm$7.6) {\bf 0.5} &
114 (107$\pm$27) {\bf 0.2}  & -\\ \hline
$\mET<50$, $\HT>200$, no Z & 11 (4.5$\pm$1.5) {\bf 72} & 
45 (51.9$\pm$6.2) {\bf 30} & -\\ \hline
$\mET<50$, $\HT<200$, no Z &123 (144$\pm$36) {\bf 0.0} &
3721 (2907$\pm$412) {\bf 0.1} & -\\ \hline
\end{tabular}
}
\caption{Number of observed, SM, and new physics ({\bf NP}) 
events. The {\bf NP} entry corresponds to 
650 GeV gluinos decaying through virtual squarks and 200 GeV
Higgsinos into 150 GeV staus.\label{multi0}}
\end{center}
\end{table}

In Table~\ref{multi0} 
we show for illustration the implications of a
LLST model with 650 GeV gluinos that are forced to decay
through virtual squark into $200$ GeV (mostly) Higgsinos,
which then go to $\tilde \tau \tau$ or $\tilde \tau \nu$
($m_{\tilde \tau}=150$ GeV).
We have 
imposed the isolation cuts and the trigger efficiencies described 
in \cite{Chatrchyan:2012ye}, and have not
included events where opposite-sign same-flavor
(OSSF) lepton pairs are within the $Z$-mass window 
($75\;{\rm GeV} < m_{\ell\ell} < 105\;{\rm GeV}$), as they combine 
larger backgrounds with a smaller signal. We obtain close to 2500 
$\tilde g\tilde g$ and $\tilde g\tilde g+{\rm jet}$ events
that after cuts translate into 530 3$\ell$ and 74 4$\ell$ 
events. Relative to the background, the 
4$\ell$ channels with 0 or 1 $\tauh$ offer the 
strongest signal, which is enough to exclude this possibility.

We find that these 4$\ell$ channels are very efficient 
to explore LLST 
SUSY. In particular, 950 GeV gluino and squark masses 
seem excluded by this analysis. In the first case
we find 97 gluino pairs that after cuts introduce 12 
$4\ell$, zero-$\tauh$
events where the SM expectation is 2.8, with similar figures for
the squarks.
In both LLST cases around 
50\% of the $4\ell$, zero-$\tauh$ events are defined by 
3 muon-like leptons plus one electron, 30\% are 4 muons, and 
20\% 2 muons and 2 electrons.

\begin{table}[ht] 
\begin{center}
{\footnotesize  
\begin{tabular}{c|c|c|c}
\hline
Selection&N$(\tauh)=0$&N$(\tauh)=1$&N$(\tauh)=2$ \\ \hline\hline
  &  obs (SM) {\bf NP} & obs (SM) {\bf NP} & obs (SM) {\bf NP} \\ \hline
4 Lepton results & \multicolumn{3}{c}{} \\ \hline
$\mET>50$, $\HT>200$, no Z & 0 (0.018$\pm$0.005) {\bf 0.5} 
& 0 (0.09$\pm$0.06) {\bf 0.9} & 0 (0.7$\pm$0.7) {\bf 0.5} \\ \hline
$\mET>50$, $\HT<200$, no Z & 1 (0.20$\pm$0.07) {\bf 1.6} &
3 (0.59$\pm$0.17) {\bf 3.0}  & 1 (1.5$\pm$0.6) {\bf 1.4}\\ \hline
$\mET<50$, $\HT>200$, no Z &0 (0.006$\pm$0.001) {\bf 0.0} &
0 (0.14$\pm$0.08) {\bf 0.0} &0 (0.25$\pm$0.07) {\bf 0.0} \\ \hline
$\mET<50$, $\HT<200$, no Z &1 (2.6$\pm$1.1) {\bf 0.1} & 
5 (3.9$\pm$1.2) {\bf 0.1}&17 (10.6$\pm$3.2) {\bf 0.0} \\ \hline
3 Lepton results & \multicolumn{3}{c}{} \\ \hline
$\mET>50$, $\HT>200$, no-OSSF &2 (1.5$\pm$0.5) {\bf 0.8} &
33 (30.4$\pm$9.7) {\bf 1.4}  &15 (13.5$\pm$2.6) {\bf 0.1} \\ \hline
$\mET>50$, $\HT<200$, no-OSSF & 7 (6.6$\pm$2.3) {\bf 1.1}  &
159 (143$\pm$37) {\bf 2.4} &82 (106$\pm$16) {\bf 0.2} \\ \hline
$\mET<50$, $\HT>200$, no-OSSF &1 (1.2$\pm$0.7) {\bf 0.0} &
16 (16.9$\pm$4.5) {\bf 0.0} & 18 (31.9$\pm$4.8) {\bf 0.0}\\ \hline
$\mET<50$, $\HT<200$, no-OSSF & 14 (11.7$\pm$3.6) {\bf 0.0} &
446 (356$\pm$55) {\bf 0.0} &1006 (1026$\pm$171) {\bf 0.0} \\ \hline
$\mET>50$, $\HT>200$, no Z &8 (5.0$\pm$1.3) {\bf 2.6}  &
16 (31.7$\pm$9.6) {\bf 1.3}  & -\\ \hline
$\mET>50$, $\HT<200$, no Z & 30 (27.0$\pm$7.6) {\bf 3.8} &
114 (107$\pm$27) {\bf 2.2}  & -\\ \hline
$\mET<50$, $\HT>200$, no Z & 11 (4.5$\pm$1.5) {\bf 0.1} & 
45 (51.9$\pm$6.2) {\bf 0.0} & -\\ \hline
$\mET<50$, $\HT<200$, no Z &123 (144$\pm$36) {\bf 0.1} &
3721 (2907$\pm$412) {\bf 0.0} & -\\ \hline
\end{tabular}
}
\caption{Number of observed, SM, and NP
events. The {\bf NP} entry corresponds to 
1070 GeV squarks decaying into 200 GeV staus through 1050 GeV
Higgsinos. \label{multi1}}
\end{center}
\end{table}

Finally, we would like to comment on another result described
in the CMS study. They find one $4\ell$ event in the 
zero-$\tauh$, no-Z, high-$\mET$, low-$\HT$ bin when the expectation
is $0.20\pm 0.07$. This observation comes together with three 
more $4\ell$ events in the $N(\tauh)=1$, no-Z, high-$\mET$, 
low-$\HT$ bin
for a background of $0.59\pm 0.17$ events (see Table~\ref{multi1}). 
Although these events are not statistically significant, we
think it is interesting to find whether LLST SUSY could explain
consistently a multilepton anomaly of this type.
The low-$\HT$ feature would be 
obtained if there is a relatively 
small mass difference between the colored particles (let us say
squarks) and the charginos/neutralinos (mostly Higgsinos), which
reduces the amount of energy going into jets. The four leptons would
result when two neutralinos go into 
$2 \tilde \tau \tau$, with
both taus decaying leptonically $\tau\to \ell \nu\bar \nu$ in the
$N(\tauh)=0$ event or one leptonically and the other one hadronically
in the 3 events with one $\tauh$.

In Table~\ref{multi1} we have taken 1070 GeV (light-flavor) 
squarks, 1050 GeV 
Higgsinos and 200 GeV staus, with the rest of SUSY particles between
1500 GeV and 2 TeV. For the quoted luminosity we obtain 92 
$\tilde q\tilde q$ pairs yielding after cuts 
a total of 8 $4\ell$ and 16 $3\ell$
events in different $\mET$, $\HT$ and $N(\tauh)$ bins. The model would
have also implications in the analysis
based on $\ST$ (the total transverse energy from jets, leptons and
$\mET$). In particular, the $4\ell$ event in the $N(\tauh)=0$ bin 
and the three events with $N(\tauh)=1$ tend to have large values of
$\ST$, as the parent particles are very heavy colored particles. 
Lower values of $\ST$ would require the direct production of the 
parent neutralino (mostly Higgsino) and masses around $\mu=400$ GeV.

\section{Summary and discussion}
SUSY has been during the past decades 
the favorite candidate to explain the 
physics above the EW scale. Unfortunately, no signs of
SUSY have been observed yet at the LHC. In this paper we
have analyzed 
how model-dependent this SUSY search has been. In particular,
we have focused on a scenario where the squarks and gluinos
created in $pp$ collisions always produce a long-lived 
stau at the end of their decay chain. We have argued that 
if their mass difference is large, most of the staus will 
be fast ($\beta > 0.8$) and will look indistinguishable from a 
muon. Instead of the large $\mET$ typical in  
neutralino LSP scenarios, these LLST models would be 
characterized by the presence of extra leptons.
We have studied how constrained they are by recent 
SS-dilepton and 
multilepton 
searches performed by ATLAS \cite{ATLAS:2012ai}
and CMS \cite{Chatrchyan:2012ye}, respectively.

We find that LLST SUSY provides signals with relatively low
SM background. The optimal search for SS dileptons would
be obtained by relaxing the cuts on $\mET$. In this sense, 
another very recent CMS analysis \cite{:2012th} of SS dileptons
at the LHC 
provides the results in each region of $E_T$ and 
$H_T$, which would allow a complete exploration of the
scenario presented here (we estimate that it could yield
bounds very similar to the ones obtained in Section 3).
 
Both in SS-dilepton and
multilepton searches the larger frequence of muons relative
to electrons
could be an interesting
observation. Notice that any model with long-lived charged
particles resulting from the decay of heavier colored ones
would imply an excess of muon-like particles, 
while the usual backgrounds (from top-quark
or vector-boson decays) are $\mu$--$e$ symmetric. 

The signature in this LLST scenario
is somewhat similar to the one
from models with broken $R$-parity and slepton decaying promptly
into lepton plus gravitino \cite{Alves:2011wf,Hanussek:2012eh}. 
Our signal, however, tends
to include less $\mET$, as the whole slepton (and not just
{\it half} of it) is visible.
Given the negative results provided so far by 
standard SUSY searches at the LHC, in order to
complete the search it seems necessary to explore 
in detail also these other SUSY possibilities.

\section*{Acknowledgments}
We would like to thank Jong Soo Kim,
Olaf Kittel and Jos\'e Santiago for valuable discussions.
This work has been partially supported by 
MINECO of Spain (FPA2010-16802
and Consolider-Ingenio {\bf Multidark} CSD2009-00064)
and by Junta de Andaluc\'{\i}a (FQM 101, FQM 03048, FQM 6552).

\end{document}